# A Novel Header Matching Algorithm for Intrusion Detection Systems

Mohammad A. Alia[1], Adnan A. Hnaif[1], Hayam K. Al-Anie[1], Khulood Abu Maria[1,] Ahmed M. Manasrah[2], M. Imran Sarwar[3]

[1] Faculty of Science and Information Technology – Al Zaytoonah University of Jordan, P.O.Box: 130 Amman (11733) Jordan
*dr.m.alia, dr.adnan_hnaif, drhayam, khulood@alzaytoonah.edu.jo*
[2] Al Yarmouk University , Irbed( 21163) – Jordan
*ahmad.a@yu.edu.jo*
[3] National Advanced IPv6 - Universiti Sains Malaysia, 11800 Penang, Malaysia
*Imran@nav6.org*

## Abstract

*The evolving necessity of the Internet increases the demand on the bandwidth. Therefore, this demand opens the doors for the hackers' community to develop new methods and techniques to gain control over networking systems. Hence, the intrusion detection systems (IDS) are insufficient to prevent/detect unauthorized access the network. Network Intrusion Detection System (NIDS) is one example that still suffers from performance degradation due the increase of the link speed in today's networks. In This paper we proposed a novel algorithm to detect the intruders, who's trying to gain access to the network using the packets header parameters such as; source/destination address, source/destination port, and protocol without the need to inspect each packet content looking for signatures/patterns. However, the "Packet Header Matching" algorithm enhances the overall speed of the matching process between the incoming packet headers against the rule set. We ran the proposed algorithm to proof the proposed concept in coping with the traffic arrival speeds and the various bandwidth demands. The achieved results were of significant enhancement of the overall performance in terms of detection speed.*

## 1. Introduction

Due to the increasing demand on the internet, the traditional boundary security systems (i.e. firewalls) that concerns swapping information between the internet and the intranet are no more efficient in providing a robust and secured network environments [1]. Hence, the firewalls only provide the first level of defense for the network that achieved by blocking unauthorized access to the network. Therefore, the need for better security systems to cover the loophole behind is also demanding. Examples of these demanding systems are: Intrusion Detection system (IDS) and Network Intrusion Detection system (NIDS).

The Intrusion Detection System (IDS) is a system that detects the intrusions in it is early stages while they are trying to steal information and/or reporting/hiding their existence to the network administrator [2]. Furthermore,

IDS is also can be defined as a process of determining the status and the stage of the attack (*attempt to attack or the attack is taking place*) [3].

However, IDS use two techniques to achieve it is main functionality in detecting and identifying intruders: *Misuse detection IDS* and *anomaly detection IDS*. The misuse detection IDS is the simplest type of detection, as it looks for a specific signature/pattern in the data part of the network traffic. This signature is known as a rule. These kinds of anomaly detection systems are widely used to monitor and identify special types of events in the network. This involves generating alerts when there are some changes to the normal system behavior under normal traffic circumstances [4]. Therefore, the misuses based detection IDS has an advantage in it is simplicity of adding known attacks to the defined rule set where on the other hand, Its disadvantage is its inability to recognize unknown attacks. In this work we will consider the performance factor of the misused detection based IDS.

On the other hand, the anomaly based IDS is for detecting intrusions and misuses by monitoring the system activities and classifying these activities as either *normal* or *anomalous*. In general, this classification is based on specific rules, rather than patterns or signatures, as well to detect any type of undefined misuse to the normal system operation [5]. However, in order to analyze the system activity from the underlying traffic, the IDS system must be able to learn how to recognize the normal system activity from the abnormal system activities that can be accomplished with the use of artificial intelligence and neural networks techniques. Even though, these kinds of IDS still require processing all types of traffic for a better detection process which is not considered as a feasible solution due to some hardware limitations such as the bus speed and the TCP/IP overhead. Therefore, some researchers have been conducted to speed up the detection process such as [6] who classifies each packet into two fields: Header and Payload (*content*). The Header contains the main information of the packet such as *source/destination address, source/destination port, and protocol*. Also they defined a rule sets that is used to determine which packets are allowed to pass to the network. However, this technique was taken from snort IDS [7] and the researchers used this technique as a design only, the researchers applied this technique on different tools, but they employ different intrusion detection algorithms.

## 2. RELATED WORKS

Snort NIDS [7] depends on the pattern matching based on the defined rule set against the captured packet to identify or recognize the intruder. Since, the number of the rules can grow over time, snort divides its rule sets into two dimensional linked lists as portrayed in Figure 1.

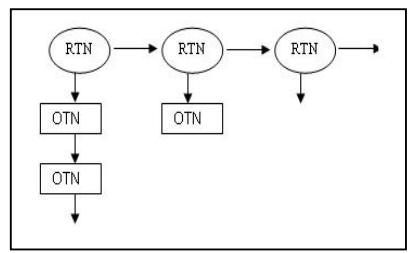

Figure 1. Structure of snort rule sets [8]

The first link list is the **Rule Tree Nodes (RTNs)** that hold the main information from each rule such as: *source/destination address, source/destination port and protocols type (TCP, ICMP, UDP)*. The second link list is the **Option Tree Nodes (OTNs)** that hold other information not exists in the RTNs such as: *TCP flags, ICMP codes and types, packet payload size and a major bottleneck for efficiency and packet contents*. These two data structures are chained together, having the RTNs are the main nodes (*chain header*) starting from left to right and the OTN's as the leave node for the RTNs (*chain Leave*). This design increased the efficiency for the intrusion detection process as it requires checking only one RTN for multi OTN.

With regards to the rule set creation, some researchers work focuses in employing the Artificial Intelligence (AI) and the Genetic Algorithm (GA) techniques in automating the rule set generation in order to enhance the overall detection performance of the IDS or the NIDS. For example, [9] utilize the GA to automatically create a rule set from the captured network traffic. These rules are stored in the rule data base that takes the following form:

$$IF \left\{ condition \right\} Then \left\{ act \right\}$$

Where, the *condition* refers to the matching case between the current network traffic flow and the current rule from the defined rule set (*e.g. source/destination address, source/destination ports, protocol, etc...),* while, *act* refers to the predefined set of actions controlled by the security policies within the organization (*e.g. reporting an alert, stopping the connection, etc...*). However, the rule set will grow exponentially as well as considering all the new rules as intrusion thus, increasing the false positive ratio.

From the other point of view, few work have been done concerning the structure of the network traffic (*packets*) to maximize the matching performance during the process of intrusion detection [10], they developed a new rule matching process for the intrusion detection systems that splits the packets header into chunks of buffers. Also, they categorized the intrusion rule set based on the packet content, into two categories: *header rule set without content, and header rule set with content*. The former category known as the *Early Filtering (EF) rule set*. The EF starts matching the packet header chunks one at the time (*packet without content*) against it is rule set. If the chunk matches with one of the EF rules, the packet will be discarded. On the other hand, if the packet is with content, the EF will match its chunks against its rule set, if it matches with one of the EF rules, the packet will be discarded, and otherwise, the packet content will be further inspected by another module. Even though the above technique, shows an interesting results towards the efficiency of the intrusion detection, but at the same time it requires more processing time for the searching and matching process with the rule set in real time.

Therefore, some of the research work focuses on the searching and matching algorithms to enhance the overall detection speed in real time as well as increasing the detection accuracy. These searching algorithms basically look for special packets within a fixed time interval. However, the special packets are set of sent/received packets from a specific IP address, specific port number, or a specific subject that might exist in the payload itself [11]. However, this gives a complete flexibility to the user to search through the range of packets that have been transmitted through the network. Therefore, [11] developed a network traffic analysis tool that mainly search for special packets within a fixed time interval as well as combining the searching results to a visualization function to interpret the result. They also design a Pre-processing function to convert the numeric and continuous data into a discrete format that can be used by different algorithms and visualization techniques. However, the above model impose a challenge on the system to cope with the packet arrival speeds in order to process all the captured traffic headers and payloads in a fast, reliable and efficient way as the whole process requires a full packet inspection.

Therefore, other researcher's focuses on enhancing the filtering process looking to speed up the matching and detection process such as [12], who propose a classification process that consists of two sequential procedures: the first procedure is to classify the incoming packets into one of the three groups: {*Accept group, Drop group, Forward group*}, where the second procedure is the action that should be taken as a filtering process to the packets grouped by the first classification procedure as follows:

**Receive packets();**
**Classify_packets(;)**
**result = filter (i);**
**action (result);**

The authors stated that, all incoming packets at a very high speed arrival (*e.g. DoS attacks*) are received by the *Receive packets ( )* function and stored in the receiving buffer for continuous processing. Therefore, they suggest executing the processing loop in shorter time by applying code optimization techniques to speed up the processing rate.

From the above example, the function filter consists of only logical and arithmetic operations. Thus, it is easy to optimize this function with various code optimization techniques. The Function *Action ( )* hosts a packet to an upper layer and send a packet to another ports and so on. This means that it is impossible to apply the optimization techniques to this part. Thus, to make the loop optimization, they divided the loop into two loops:

**For (i = 0; I < n_packets; i ++)**
  **{**
    **result[ i ] = filter (i);**
  **}**
**For (i = 0; I < n_packets; i ++)**
 **{**
  **action (result [i]);**
 **}**

The first loop processes multiple consecutive packets together utilizing a software pipelined procedure. This pipelining procedure is considered a highly sophisticated aggressive instruction scheduling technique for loops that occupy the whole CPU time and thus waste it is processing time for other functions to be executed.

## 3. PHM Design

The current packets header detection engines use any exact string matching algorithms to detect the intrusions. These string matching algorithms deal with decimal, hexadecimal, and characters values. The PHM algorithm deals with binary values, because in the proposed methodology, the headers rule set is converted into 8 weights. On the other hand, PHM algorithm saves memory, because only 12-bits are reserved from the total memory size.

The rule sets are classified into two sections: headers rule set, and payloads rule set. Headers rule set contains the main information of the packet such as: source/destination address, source/destination port, and protocol. Payloads rule set contains the real data of the packet. Each rule represents one different type of the intruders. To commensurate the headers rule set with the proposed design and methodology, the headers rule set is converted into binary as depicted by Figure 2.

| Rule ID | Rule (SA, SP, DA, DP, and PROT) |
|---------|-------------------------------|
| 1 | 111010101011111110000000010101010000001010101010101010...10 |
| 2 | 1000010101010101010000000110111111110010101010101111...01 |
| 3 | 1100000001111101010101111101010011111100101011111011...11 |
| 4 | 1010101010101010101111111101010101000000000111111111011...00 |
| . | . |
| . | . |
| . | . |
| . |  |
| N | 111111111111111111111100000000000000000001101010101010101010...10 |

Figure 2. Design of the header rule sets

The aim of the PHM algorithm is to enhance the speed of the detection engine for packets header of any NIDS. The PHM algorithm consists of the following steps:

1- Converting headers rule sets into weight, as the energy function is required.
2- Matching process.
3- Learning process that can be used to increase the performance of PHM algorithm.

### 3.1. Converting The Rule Set Into Weight

In order to evaluate the matching process between two values, we will apply the energy Function [13] that requires the conversion of the rule set into weights. Therefore, we divided each header rule set into 3-bit at the time, then we converted each 3-bit ($2^3 = 8$ possibilities) into a symmetric matrix of weight. However, to avoid memory exhaustion, this matrix of weights requires only 24 bits of memory ($3 \ x \ 8 \ = 24$).

On the other hand, the proposed algorithm uses a neural network with a multi-connect architecture [14] to learn the header rule sets that is described as an associative memory and a single-layer neural network. The neural network with the multi-connect architecture will learn the set of pattern pairs *(associations)* and store the patterns set in the memory. These memory values will be activated during the retrieval process with the key pattern that contains a portion of the information related to a particular stored pattern set. The learning process creates eight learning weights for all possibilities of the patterns matrices (of the size 3*3) as portrayed in Figure 3.

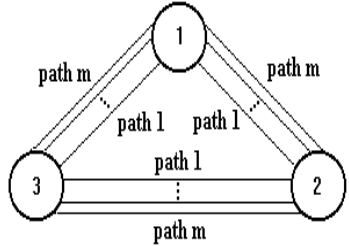

Figure 3. NN with multi-connect architecture [14]

As shown in Figure 3, each path represents one learning weight (m, $\forall\ 0 \le m < 8$), thus, a matrix of these paths by the size of the net path value $\sum_{i=0}^{7} m_i$ is constructed. The learning process will be a one-time only process.

For instance, suppose that we have the header rule in a binary format as follows:

$$R[i], \forall R[i] \in [0,1], i \in [0,8)$$

| 1 | 1 | 0 | 1 | 1 | 1 | 0 | 0 | 1 | 0 | 1 | 0 | … |
|---|---|---|---|---|---|---|---|---|---|---|---|---|

To convert the header rule sets into a matrix of weight we will apply the Hopfield nets [15] into the rule set. The Hopfield nets requires the units to binary threshold units, i.e. the units only take on two different values for their states and the value is determined by whether or not the units' input exceeds their threshold. Hopfield nets can either have units that take on values of 1 or -1, or units that take on values of 1 or 0. We choose 1 or -1 value because we will be applying the "energy function" [13] on the result in a later stage. The energy function should results in – 3 values to indicate stability. However, the conversion process is discussed as follows:

**A.** Convert each 0 into -1 as follow:

$$R[i] = \begin{cases} R[i], & R[i] \ne 0 \\ -1, & R[i] = 0 \end{cases}$$

| 1 | 1 | -1 | 1 | 1 | 1 | -1 | -1 | 1 | -1 | 1 | -1 | … |
|---|---|----|---|---|---|----|----|---|----|---|----|---|

**B.** From the first three bits (from the left) assemble column vector $(r)_{i,j}$ matrix and multiply it by $(r^T)_{i,j}$ recursively to produce the matrix $(S)_{i,j}$ , as follows:

$$(S)_{i,j} = (r)_{i,j} \times (r^T)_{i,j} = \sum_{i=j}^{3} (r)_{i,j} (r)_{j,i},$$

$$1 \leq i \leq 3, \quad 1 \leq j \leq 3,$$

For instance,

$$(r)_{i,j} = \begin{bmatrix} 1 \\ 1 \\ -1 \end{bmatrix},$$

$$(r^T)_{i,j} = \begin{bmatrix} 1 & 1 & -1 \end{bmatrix}$$

Therefore,

$$(S)_{i,j} = (r)_{i,j} \times (r^T)_{i,j} = \begin{bmatrix} 1 & 1 & -1 \\ 1 & 1 & -1 \\ -1 & -1 & 1 \end{bmatrix}$$

and so on.

The connections in a Hopfield net typically have the following restrictions:

- No unit has a connection with itself $(S)_{i,i} = 0, \forall i \in \{1,2,...n\}$.

- Connections are symmetric, and thus, $(S)_{i,j} = (S^T)_{i,j}$

**C.  Zero diagonal:**

Since there is no unit has a connection with itself, then

$$(S)_{i,j} = 0, \quad \forall i = j$$

$$= \begin{bmatrix} 0 & 1 & -1 \\ 1 & 0 & -1 \\ -1 & -1 & 0 \end{bmatrix}$$

As a result, the three bits located above the diagonal $(s)_{i,j} \ \forall j > i$, represents the weight for the first three bits (1 -1 -1) from the rule set. and Since we represent each 0 value with -1, a backward substitution (0 instead of -1) will result in decimal value equal to 4 (1 0 0). Therefore, we can replace the 3 bits with 4.

**D.**  If the following 3 bits i.e. (1 1 1) weight is calculated previously, the same weight value will be assigned to them, otherwise, the weight calculation process (as in B) will be performed again. There is no need to save another matrix, because it is already exist in the memory, because only 8 matrices of 3x3 need to be saved into memory.

There are some cases where conflict between two dimensional arrays $(r)_{i,j}$ weight calculation may occur (i.e. $\begin{bmatrix} 0 & 0 & 0 \end{bmatrix}$ and $\begin{bmatrix} 1 & 1 & 1 \end{bmatrix}$). These two different arrays will have the same results when they are multiplied by their $(r^T)_{i,j}$. Figure 4 depicts the same results matrices for values in identical colored boxes.

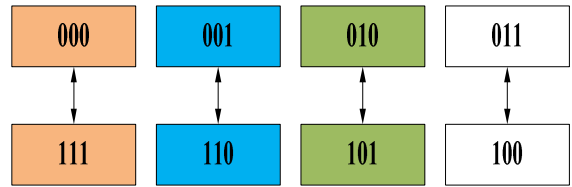

Figure 4. The same results matrices for values in identical colored boxes

Consequently, 24 bit of memory can be minimized to only 12 bit as depicted in Figure 5.

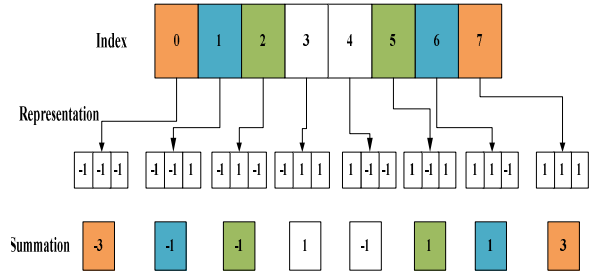

Figure 5. Storing process

As shown in Figure 5, it can be noted that each two same results matrices have difference in their summation index. Thus, the problem of inconsistency is solved.

Table 1 depicts the header rule set after the conversion into weight, where each row represents one header rule from the original header rule set. For efficient processing, and especially dealing with intrusions, the calculated weights will be rearranging in a descending order having an index of each group in the first column.

Table 1. Weight of the headers rule set

| Group # | $W_1$ | $W_2$ | $W_3$ | $W_4$ | $W_5$ | $W_6$ | $W_7$ | $W_8$ |
|---|---|---|---|---|---|---|---|---|
| 7 | 7 | 7 | 6 | 4 | 2 | 1 | 0 | 7 |
|  | 7 | 6 | 5 | 1 | 5 | 3 | 7 | 2 |
|  | 7 | 6 | 4 | 6 | 3 | 5 | 4 | 0 |
|  | 7 | 0 | 7 | 3 | 4 | 5 | 6 | 2 |
| 6 | 6 | 4 | 5 | 3 | 6 | 7 | 5 | 1 |
|  | 6 | 3 | 6 | 2 | 7 | 3 | 5 | 1 |
| 5 | 5 | 7 | 3 | 5 | 2 | 5 | 3 | 1 |
| 4 | 4 | 7 | 1 | 6 | 3 | 5 | 4 | 2 |
|  | 4 | 5 | 3 | 2 | 6 | 7 | 4 | 3 |
| 3 | 3 | 7 | 4 | 3 | 2 | 1 | 5 | 4 |
|  | 3 | 4 | 2 | 1 | 7 | 4 | 2 | 5 |
| 2 | 2 | 6 | 3 | 4 | 1 | 0 | 4 | 3 |
| 1 | 1 | 6 | 4 | 7 | 2 | 5 | 4 | 3 |
|  | 1 | 1 | 2 | 3 | 4 | 4 | 4 | 4 |
| 0 | 0 | 5 | 3 | 2 | 5 | 4 | 3 | 7 |
|  | 0 | 2 | 5 | 7 | 3 | 6 | 1 | 3 |

Where W is: weight for each 3-bit

### 3.2. Matching Phase

Most of the previous work exists in the literature, uses an exact string matching algorithms such as Boyer-Moore and Horspool algorithm [16] to find a specific pattern in a given text. Therefore, in this paper we applied a different algorithm called PHM algorithm to match the header rule set with the incoming packet header. This algorithm works as follows:

a- For every incoming packet header, convert each 0 into -1.
b- From left to right, take the first three bits.
c- Apply "Lyapunov function" or "Energy function" [13]:

$$EF = LF = -\frac{1}{2} \times (\sum_{i=1}^{n} \sum_{j=1}^{n} x_i x_j w_{ij}) = -3$$

Where,

$n$: the number of elements in the each matrix $(r)_{i,j}$ (which is equal to 3),

$w_{ij}$: is the calculated weight from the 3-bit in $(r)_{i,j}$.

For instance, suppose the incoming packet header as follows:

| 1 | 0 | 0 | 1 | 0 | 1 | 1 | 0 | 1 | 0 | 0 | ... |
|---|---|---|---|---|---|---|---|---|---|---|-----|

From step **a**, convert each 0 into -1.

| 1 | -1 | -1 | 1 | -1 | 1 | 1 | -1 | 1 | -1 | -1 | ... |
|---|----|----|---|----|---|---|----|---|----|----|-----|

From step **b**, the first three incoming bits are kept in $(r)_{i,j}$ to be multiplied by it is $(r^T)_{i,j}$ to construct the $(S)_{i,j}$ matrix as follows:

$$(S)_{i,j} = (r)_{i,j} \times (r^T)_{i,j} = \sum_{i=j}^{3} (r)_{i,j} (r)_{j,i} ,$$

$$1 \le i \le 3, \quad 1 \le j \le 3,$$

I =     1     2     3

| 1 | -1 | -1 |
|---|----|----|

J =     1     2     3

The matching process starts by applying the "energy function" on the calculated weights from the incoming packet header. The matching process will start the matching with each weight group from the rule set weight matrix (starting from group-7 to group-1). If there is a match (energy function returns -3) the algorithm identifies the matching group from the header rule set to search within. Otherwise, the energy function will be applied to the next group (i.e. group-6) and so on.

The matching process continues the matching process until group-1, if there are no matches yet; the matching process will assume that there is a matching with the group-0 (zero). This matching considered a trigger to evaluate the second three bits from the incoming packet header by applying the energy function on them in group-0.

For instance, with Group-7 the weight matrix will be:

$$\begin{bmatrix} 0 & 1 & 1 \\ 1 & 0 & 1 \\ 1 & 1 & 0 \end{bmatrix}$$

and the incoming packet header is $\begin{bmatrix} 1 & -1 & -1 \end{bmatrix}$,

Therefore, EF= 1, and since EF $\neq -3$, that there is no rule from the rule set matches with the header. As a result, the matching will proceed to group-6, and so on until group-4 where the weight matrix is:

$$\begin{bmatrix} 0 & -1 & -1 \\ -1 & 0 & 1 \\ -1 & 1 & 0 \end{bmatrix}$$

And the incoming packet header is $\begin{bmatrix} 1 & -1 & -1 \end{bmatrix}$, therefore, EF = -3, which indicates a match with the rule set under group-4, and the search in others group should not continues. In this case, the second three bits from incoming packet header will be matched with the second weight from group-4, until all the incoming packet header matches to one rule from group-4, otherwise, until any mismatch; which indicate no possibilities to find the incoming packet header within the rule set.

### 3.3. The Learning Process

Since all incoming packet header matched with group-4, therefore, the three bits $\begin{bmatrix} 1 & -1 & -1 \end{bmatrix}$ is always matching with group- 4. Thus, a matching link list can be created that contains the matched bits and the group number to where the bits are matched as depicted in Figure 6.

| Three Bits | Group |
|------------|-------|
| 1 -1 -1 | 4 |

Figure 6. Matching Link List

This is a very important step towards speeding up the searching and matching process. Therefore, in the next matching process, the algorithm will first check the current 3 bits in the link list (no need to apply the energy function), because the algorithm learned the system in advance for the expected three bits with its corresponding group, otherwise, the energy function will be applied until a matched group found the result will be added to the link list. This process continuous until the 8 possibilities is included in the link list.

## 4. PERFORMANCE ANALYSIS OF THE PROPOSED ALGORITHM

We compared the performance of the proposed PHM algorithm against the well known SNORT Algorithm. Table 2 shows the performance for both approaches. Both algorithms were coded in C++ on a computer with 1.6 GHz Intel® M Pentium processor and 1 GB RAM.

Table 2. Performance evaluation between PHM and SNORT algorithms

| # of packets | PHM | SNORT |
|---|---|---|
| 450 | 0.28 | 0.60 |
| 500 | 0.28 | 0.67 |
| 550 | 0.37 | 0.73 |
| 600 | 0.38 | 0.85 |
| 650 | 0.42 | 0.84 |
| 700 | 0.58 | 0.91 |
| 900 | 0.82 | 1.23 |
| 1200 | 1.14 | 1.64 |
| 2500 | 2.33 | 3.22 |
| 3000 | 3.40 | 3.80 |
| 4500 | 4.71 | 5.93 |
| 5500 | 5.83 | 7.20 |
| 8000 | 7.82 | 10.47 |
| 10000 | 10.78 | 13.08 |

## 4.1. Evaluation Process

Based on the hardware and software architecture, we read 450 to 10000 packets from the file. The header rule sets had a size of 50 KB. In order to evaluate the effectiveness of PHM algorithm, a comparison is made with SNORT which uses Boyer-Moore algorithm. The comparison results can be viewed in Figure 7.

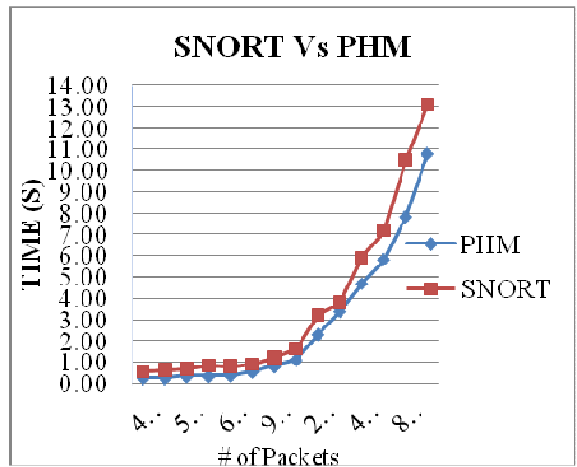

Figure 7. Overall time comparison between PHM and SNORT algorithms

From Figure 7, it can be seen that the performance of PHM algorithm and Boyer-Moore algorithm are very close for the first 1200 packets. After this point, it seems that the PHM algorithm has learnt all the header rule sets. However, it is clear that the PHM algorithm is faster after 1200 packets as compared to Boyer-Moore algorithm. This is a very encouraging enhancement. The achieved improvement was between 10% - 58%.

## 5. CONCLUSION

In this work we have presented a novel algorithm to speed up the detection engine for packets header. We divided the work into three phases, namely: classification of rule sets, matching process, and the learning process. As the result, the proposed Packet Header Matching (PHM) algorithm is more efficient than SNORT algorithm. It requires a much lower cost of execution time and performs at a high level of detection compared to the exiting SNORT algorithm.

## ACKNOWLEDGMENT

The researchers would like to thank Al- Zaytoonah University of Jordan for supporting this study.

# AUTHORS

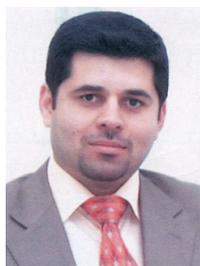

**Dr. Mohammad Alia** is an Assistance professor at the computer information systems department, Faculty of science Computer and information technology, Al Zaytoonah University of Jordan. He received the B.Sc. degree in Science from the Alzaytoonah University, Jordan, in 1999. He obtained his Ph.D. degree in Computer Science from University Science of Malaysia, in 2008. During 2000 until 2004, he worked at Al-Zaytoonah University of Jordan as an instructor of Computer sciences and Information Technology. Then, he worked as a lecturer at Al-Quds University in Saudi Arabia from 2004 - 2005. Currently he is working as a Chair of Computer Information Systems dept. at Al Zaytoonah University of Jordan. His research interests are in the field of Cryptography, and Network security.

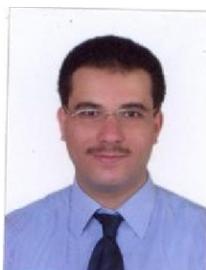

**Dr. Adnan Hnaif** is an Assistance professor at the computer information systems department, Faculty of Science Computer and information technology, Al Zaytoonah University of Jordan. Dr. Hnaif received his PhD degree in Computer Science from University Sains Malaysia - National Advanced IPv6 Centre and Excellence (NAV6) in 2010. He received his MSc degree of Computer Science from department of Computer Science- Alneelain University in 2003, and obtained his Bachelor degree of Computer Science from the department of Computer Science, Mu'tah University in 1999/2000. His researches focus on the network security and parallel processing.

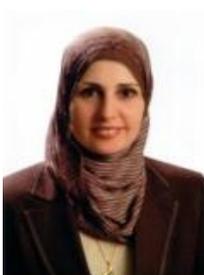 **Dr. Hayam Al-Anie** is an Assistance professor at the computer information systems department, Faculty of Science Computer and Information Technology, Al Zaytoonah University of Jordan. She received the B.Sc. degree in Computer Science from the University of Technology, Iraq, in 1988. She obtained her Ph.D. degree in Computer Science from University of Technology, Iraq in 2004. During 1988 until 1996, she worked at University of Baghdad, College of Administration and Economic, as programmer assistance. Then she worked as an instructor of Computer Science and Information Technology at University of Baghdad, College of Administration and Economic from 1996-2004. Currently she is working as an Assistance professor at Al Zaytoonah University of Jordan. Her research interests are in the field of Cryptography, and Software Engineering.

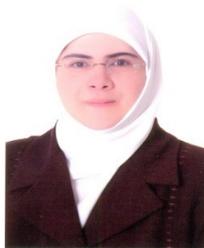 **Dr. Khulood Abu Maria** is an Assistance professor at the computer information systems department, Faculty of science Computer and information technology, Al Zaytoonah University of Jordan. She received the B.Sc. degree in Science from Mut'ah University, Jordan, in 1992. She obtained her Ph.D. degree in Computer Information System from Arab Academy for Banking and Financial Sciences, Jordan, in 2008. During 1992 until 2006, she worked at Petra Engineering Industries Co. as analyst, programmer, Network Administrator, Quality Assurance and IT manager. Then, she worked as a part-time instructor at AL-ISRA Private University, Jordan from 2008 - 2009. Currently she is working as an instructor at Al Zaytoonah University of Jordan. Her research interest is in the fields of AI, Agent-Based Application, Information System, Software Engineering, E-applications, m-commerce, Security, and Network.

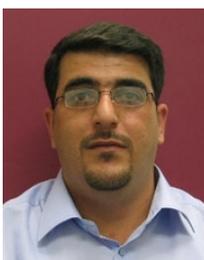 **Dr. Ahmed** obtained his Bachelor of Computer Science from Mu'tah University, al Karak, Jordan in 2002. He obtained his Master of Computer Science and doctorate from Universiti Sains Malaysia in 2005 and 2009 respectively, he was the Deputy Director (Research and Innovation) and the Head of iNetmon project at the National Advanced IPv6 Centre of Excellence (NAV6) in Universiti Sains Malaysia. He started his career as a web developer at Telaterra LLC from 2003 until 2004. Between 2005 until 2008, Dr. Ahmed worked as senior research officer in NRG, Universiti Sains Malaysia. Upon his graduation, he worked as a senior Vice President in iNetmon Sdn. Bhd. from 2005-2008. Currently he is working as an Assistance professor at Al Yarmouk University. His research interests are in the field of network monitoring.

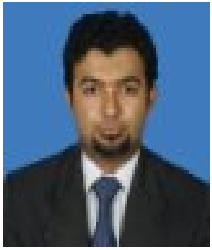

**Mr. Muhammad Imran Sarwar** is currently Ph. D fellow and researcher in National Advance IPv6 Centre (NAv6), Universiti Sains Malaysia. He received his M.Sc degree from School of Computer Sciences, Universiti Sains Malaysia in 2009. His research interest are in IPv6 multicast networks, wireless mesh networks, WiMAX networks, network security, location-based services etc.